\documentclass[pra,showpacs,onecolumn,byrevtex]{revtex4}
\usepackage{graphicx}
\usepackage{subfigure}
\usepackage{dcolumn}
\usepackage{amsmath}
\usepackage{epsfig}
\usepackage{color}

\begin{document}
\title{Bright six-partite continuous variable entanglement using cascaded four-wave mixing processes in a four-level atomic system}
\author{Guangqiang He$^1$}
\email{gqhe@sjtu.edu.cn}

\author{Xufei Wu$^1$$^,$$^2$}
\email{arnoxufei724@sjtu.edu.cn}

\author{Yi Gu$^1$}

\author{Guihua Zeng}

\affiliation{$1$ State Key Laboratory of Advanced Optical Communication Systems and Networks, Department of Electronic Engineering,
              Shanghai Jiao Tong University, Shanghai 200240, China\\$2$ Department of Physics, Shanghai Jiao Tong University,  Shanghai 200240, China}

\date{September 4, 2011}

\begin{abstract}
We theoretically show that bright six-partite continuous-variable entanglement can be generated using cascaded four-wave mixing effects of third-order nonlinearity atomic systems above threshold. The six-partite continuous-variable entanglement among the six cavity fields with different frequencies is analyzed by applying optimized inseparability criteria proposed by Van Loock and Furusawa. 

\end{abstract}

\maketitle

\section{Introduction}
Quantum entanglement is one of the most mysterious quantum phenomena and has become an indispensable resource
for quantum communication and quantum computation. In the past few years, there has been much effort devoted to the study of continuous-variable (CV) quantum communication \cite{vanloock2000,Yonezawa2004}, of which the basis is the multi-partite entanglement. Multi-partite CV entanglement is essential and necessary for measurement-based quantum computation and controlled quantum teleportation. It was predicted theretically and demonstrated experimentally that multi-color CV entanglement can be produced by nondegenerate optical parametric oscillator (OPO), optical superlattice and second-order nonlinearity media \cite{Leng2009,Midgley2010,Coelho2009,Villar2006}. Alternatively, CV entanglement can also be produced in coherent atomic systems, whose entangled beams have narrower linewidths and longer coherence times; in particular, they can be used for free-space quantum teleportation \cite{freespaceteleportation2010,freespaceteleportation12010,freespaceteleportation22010}, such as, from a satellite to a ground station. Such entanglement using four-wave mixing (FWM) has been predicted by Ref. \cite{Yu2011,Tan2010,Boyer2008}.
All those results above endow us the feasibility and necessity to realize our scheme. Based on those processes, we  extend the scheme to generate six-partite CV entanglement by intracavity four-wave-mixing cascaded with double such generations in a four-level atomic system. And we theoretically demonstrate such possibility of multy-partite entanglement using third-order nonlinearity according to the inseparability criterion for multipartite CV entanglement proposed by van Loock and Furusawa.

This paper is arranged as follows. The physical model and its hamiltonian are discussed in Sec. II. Sec. III gives the equations of motion of the system and Sec. IV provides a linearized fluctuation analysis to calculate the measurable output fluctuation spectra. These output spectra are further discussed in Sec. V and they demonstrate the violation of the optimized VLF criteria and some entanglement characteristics. At last, we draw some brief conclusions in Sec. VI.

\section{Physical Model and its Hamiltonian}
We consider an ensemble of independent four-level atoms inside a resonant optical cavity. The proposed experimental setup for this scheme is shown in Fig. 1(a). A vapor cell containing a four-level atomic system is placed inside a Fabry P'erot cavity, using lasers whose frequency are $\omega_{p1}$ and $\omega_{p2}$ as pumps, where the virtual energy level diagram of three cascaded FWM effects are shown in Fig. 1(b). Those generated six beams are transmitted out through the coupling mirror 2 with partial tramittion coefficient, then being splitted by the prism. Spatially separated beams can be measured using method as Ref. \cite{Villar2005} to determine the entanglement relationship.

As shown in Fig. 1, the lower levels, $|1\rangle$, $|2\rangle$ and $|3\rangle$, of the atoms are coupled to the excited level $|4\rangle$ by two driving lasers of frequencies $\omega_{p1}$, $\omega_{p2}$. There are three FWM processes which are indicated in  Fig. 1(b), that $1\uparrow$ and $1\downarrow$ compose the first FWM process, where $1\uparrow$ implies $\omega_{p1}$ and $\omega_{p2}$ stimulated from $|1\rangle$ and $|2\rangle$ to virtual levels, $1\downarrow$ implies $\omega_{s1}$ and $\omega_{i2}$ falling from virtual levels to $|2\rangle$ and $|1\rangle$, respectively. Similar are to $2\uparrow$, $2\downarrow$, $3\uparrow$ and $3\downarrow$. And we suppose all the virtual levels are detuned with $|4\rangle$.


\begin{figure}
\centering
\subfigure[]{
\label{fig:subfig:a} 
\includegraphics[width=0.4\textwidth]{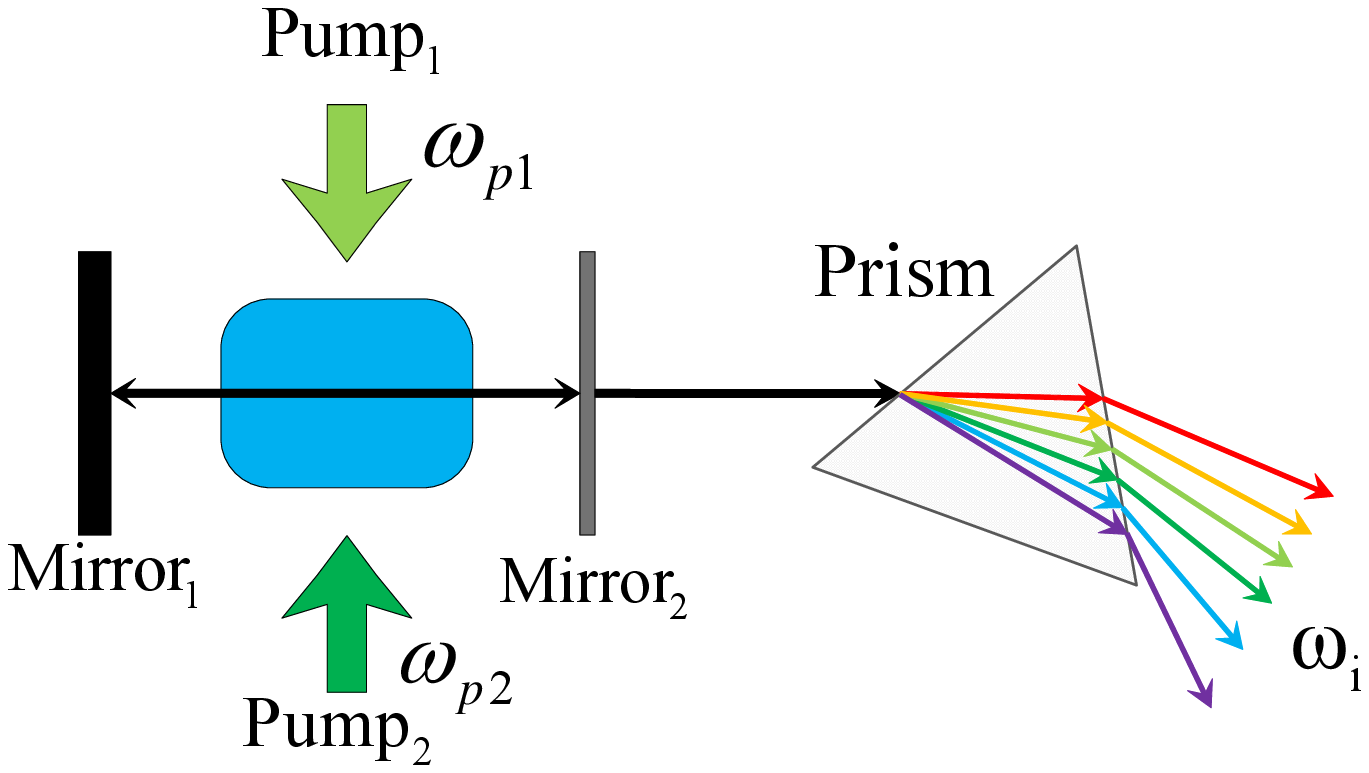}}
\hspace{0.5in}
\subfigure[]{
\label{fig:subfig:a} 
\includegraphics[width=0.40\textwidth]{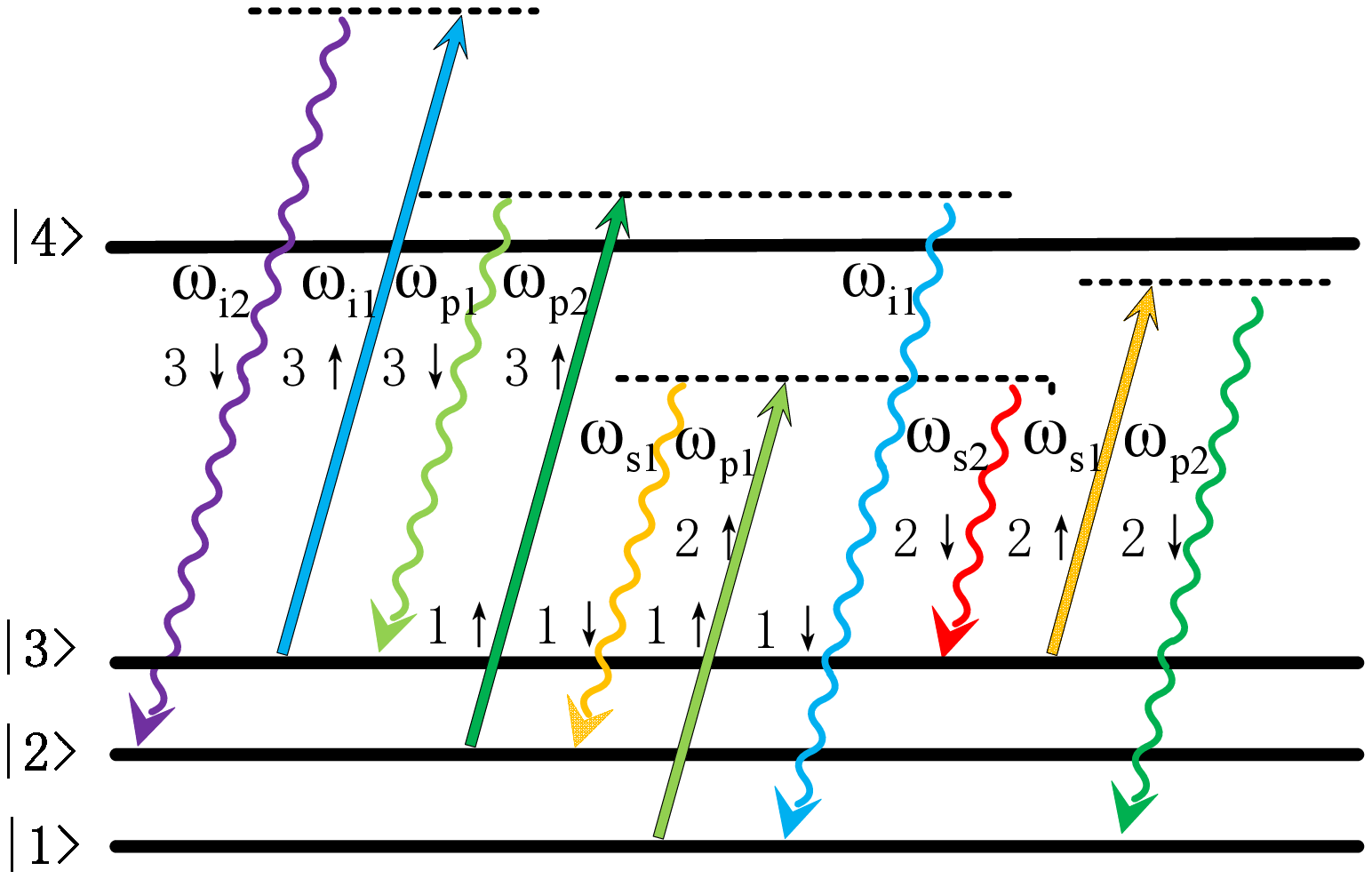}}
\hspace{1in}
\subfigure[]{
\label{fig:subfig:b} 
\includegraphics[width=0.40\textwidth]{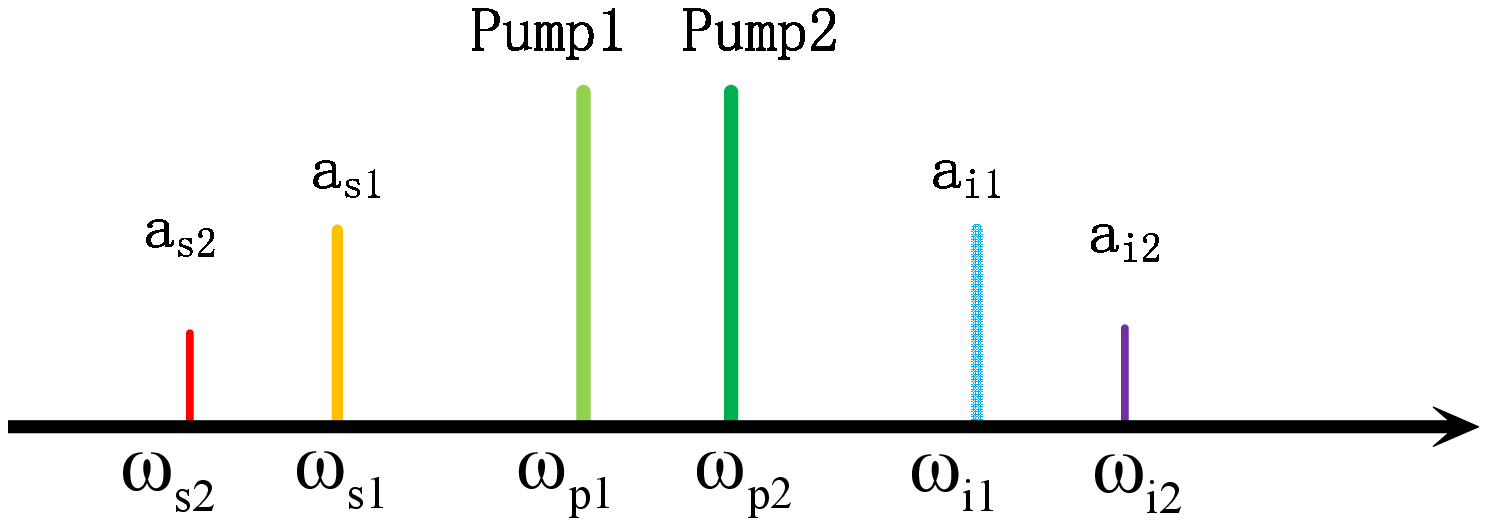}}
\caption{(Color online) (a) Experimental setup for the proposed scheme. (b) Relevant energy level diagram of the four-level atomic system. This involves three FWM processes. (c) Frequencies of the pumps and cavity fields. All the processes and photons are marked with different colors for simplicity.}
\label{fig:subfig} 
\end{figure}

In present scheme, we only concentrate on the entanglement characteristics among optical fields instead of these between atoms and optical fields. So the interaction Hamiltonian of the three cascaded FWM processes could be written as
\begin{equation}
H_{int} = i\hbar [k_{1} \hat{a}_{p1} \hat{a}_{p2} \hat{a}^{\dagger}_{s1} \hat{a}^{\dagger}_{i1} +
k_{2} \hat{a}_{s1} \hat{a}_{p1} \hat{a}^{\dagger}_{s2} \hat{a}^{\dagger}_{p2} +
k_{3} \hat{a}_{i1} \hat{a}_{p2} \hat{a}^{\dagger}_{p1} \hat{a}^{\dagger}_{i2}] + \textup{H.c.} \; ,
\end{equation}
where $k_{i}$ (i=1,2,3) are the dimensionless nonlinear coupling coefficients and are commonly taken to be a real number, $a_{i}$ and $a_{i}^{\dagger}$ are corresponding to the light field generation and annihilation operators repectively.

Considering the driving fields in the cavity, the Hamiltonian for the pump beams is given by
\begin{equation}
H_{pump} = i\hbar [\epsilon_{1} \hat{a}^{\dagger}_{p1} + \epsilon_{2} \hat{a}^{\dagger}_{p2} ] + \textup{H.c.} \; ,
\end{equation}
where $\epsilon_{i}$ (i=1,2) is the classical pump and coupling laser amplitude. The loss of the $i$th mode \cite{walls1994} in the cavity can be written as
\begin{equation}
L_{i} \hat{\rho} = \gamma_{i} [2\hat{a}_{i} \hat{\rho} \hat{a}^{\dagger}_{i} - \hat{a}^{\dagger}_{i} \hat{a}_{i} \hat{\rho} - \hat{\rho} \hat{a}^{\dagger}_{i} \hat{a}_{i} ] \; ,
\end{equation}
where $\gamma_{i}$ stand for the damping rates for the corresponding cavity modes which are related to the amplitude transmission coefficients.

\section{Equations of Motion for the Full Hamiltonian}

We now consider the full physical system, where the four-level atomic system is contained inside a pumped resonant Fabry-P'erot cavity. The master equation for the six cavity modes is
\begin{equation}
\frac{\partial \hat{\rho}}{\partial t} = -\frac{i}{\hbar}[H_{pump}+H_{int},\hat{\rho}]+\sum_{i=1}^{6}L_{i}\hat{\rho} \; .
\end{equation}

The equations of motion for the six cavity modes can be obtained by solving the Fokker-Planck equation in the P-representation \cite{walls1994} as
\begin{equation}
\frac{\partial}{\partial t} \mathbf{\alpha} = \mathbf{F} +\mathbf{B\eta} \; ,
\end{equation}
where $\mathbf{\alpha}$ can be expressed as $ [\alpha_{p2},\alpha_{p1},\alpha_{i1},\alpha_{s1},\alpha_{i2},
\alpha_{s2},\alpha_{p2}^{*},\alpha_{p1}^{*},\alpha_{i1}^{*},\alpha_{s1}^{*},\alpha_{i2}^{*},\alpha_{s2}^{*}]^{\mathbf{T}} $ and
$ \mathbf{\eta}$ be expressed as $ [\eta_{1}(t),\eta_{2}(t) ... \eta_{11}(t),\eta_{12}(t)]^{\mathbf{T}} $, where elements in the vector $\eta$ are real noise terms and have $\langle\eta_{i}(t)\rangle=0$,  $\langle\eta_{i}(t)\eta_{j}(t')\rangle=\delta_{ij}\delta(t-t')$.

And $\mathbf{F}$ is the main part of system's evolution, as the form of $\mathbf{F}=[\mathbf{f},\mathbf{f}^{*}]$, where
\begin{equation}
\mathbf{f} =
\left( \begin{array}{ccc}
\epsilon_{2}-\gamma_{p2}\alpha_{p2}-k_{1}\alpha_{p1}^{*}\alpha_{s1}\alpha_{i1}-k_{2}\alpha_{i1}^{*}\alpha_{p1}\alpha_{i2}+k_{3}\alpha_{s2}^{*}\alpha_{s1}\alpha_{p1}\\
\epsilon_{1}-\gamma_{p1}\alpha_{p1}-k_{1}\alpha_{p2}^{*}\alpha_{s1}\alpha_{i1}-k_{3}\alpha_{s1}^{*}\alpha_{p2}\alpha_{s2}+k_{2}\alpha_{i2}^{*}\alpha_{i1}\alpha_{p2}\\
-\gamma_{i1}\alpha_{i1}+k_{1}\alpha_{s1}^{*}\alpha_{p1}\alpha_{p2}-k_{2}\alpha_{p2}^{*}\alpha_{p1}\alpha_{i2}\\
-\gamma_{s1}\alpha_{s1}+k_{1}\alpha_{i1}^{*}\alpha_{p1}\alpha_{p2}-k_{3}\alpha_{p1}^{*}\alpha_{p2}\alpha_{s2}\\
-\gamma_{i2}\alpha_{i2}+k_{2}\alpha_{p1}^{*}\alpha_{p2}\alpha_{i1}\\
-\gamma_{s2}\alpha_{s2}+k_{3}\alpha_{p2}^{*}\alpha_{p1}\alpha_{s1}\\
\end{array} \right)
\;.
\end{equation}

 $\mathbf{B}$ contains the coefficients of the noise terms and is not calculated here. $\mathbf{B}$  can be obtained by factoring the diffusion matrix  $\mathbf{D}$  according to $\mathbf{D}=\mathbf{B}\mathbf{B^{T}} $ .

We express D as
\begin{equation}
\mathbf{D} =
\left( \begin{array}{ccc}
\mathbf{d} & \mathbf{0} \\
\mathbf{0} & \mathbf{d^{*}}\\
\end{array} \right) \; ,
\end{equation}
where $\mathbf{0}$ is a 6$\times$6 zero matrix and the nonzero block is given by
\begin{equation}
\mathbf{d} =
\left( \begin{array}{cccccc}
0 & -k_{1}\alpha_{s1}\alpha_{i1}   & -k_{2}\alpha_{i2}\alpha_{p1}  &  0 & 0 & k_{3}\alpha_{s1}\alpha_{p1} \\
-k_{1}\alpha_{s1}\alpha_{i1}   & 0 & 0  & -k_{3}\alpha_{s2}\alpha_{p2} & k_{2}\alpha_{i1}\alpha_{p2} & 0  \\
-k_{2}\alpha_{i2}\alpha_{p1}   & 0 & 0  & k_{1}\alpha_{p1}\alpha_{p2} & 0 & 0 \\
 0   & -k_{3}\alpha_{s2}\alpha_{p2}    & k_{1}\alpha_{p1}\alpha_{p2}  & 0 & 0 & 0 \\
 0   &  k_{2}\alpha_{i1}\alpha_{p2}  & 0 & 0 & 0 & 0 \\
k_{3}\alpha_{s1}\alpha_{p1}  &  0 & 0 & 0 & 0 & 0 \\
\end{array} \right) \; ,
\end{equation}

\section{Stability Analysis }

Now we conduct the stability analysis of the system in order to analyze the output spectral correlations.
By neglecting the noise terms in Eq. (5), we obtain a set of classical equations for the mean values:

and from these we can obtain steady-state solutions.
\begin{equation}
 \begin{split}
\frac{\partial}{\partial t} \alpha_{p2} = \epsilon_{2}-\gamma_{p2}\alpha_{p2}-k_{1}\alpha_{p1}^{*}&\alpha_{s1}\alpha_{i1}-k_{2}\alpha_{i1}^{*}\alpha_{p1}\alpha_{i2}+k_{3}\alpha_{s2}^{*}\alpha_{s1}\alpha_{p1}
 \; ,\\
\frac{\partial}{\partial t} \alpha_{p1} = \epsilon_{1}-\gamma_{p1}\alpha_{p1}-k_{1}\alpha_{p2}^{*}&\alpha_{s1}\alpha_{i1}-k_{3}\alpha_{s1}^{*}\alpha_{p2}\alpha_{s2}+k_{2}\alpha_{i2}^{*}\alpha_{i1}\alpha_{p2}
 \; ,\\
\frac{\partial}{\partial t} \alpha_{s1} = -\gamma_{i1}\alpha_{i1}+&k_{1}\alpha_{s1}^{*}\alpha_{p1}\alpha_{p2}-k_{2}\alpha_{p2}^{*}\alpha_{p1}\alpha_{i2}
 \; ,\\
\frac{\partial}{\partial t} \alpha_{i1} = -\gamma_{s1}\alpha_{s1}+&k_{1}\alpha_{i1}^{*}\alpha_{p1}\alpha_{p2}-k_{3}\alpha_{p1}^{*}\alpha_{p2}\alpha_{s2}
 \; ,\\
\frac{\partial}{\partial t} \alpha_{s2} = -\gamma_{i2}&\alpha_{i2}+k_{2}\alpha_{p1}^{*}\alpha_{p2}\alpha_{i1}
 \; ,\\
\frac{\partial}{\partial t} \alpha_{i2} = -\gamma_{s2}&\alpha_{s2}+k_{3}\alpha_{p2}^{*}\alpha_{p1}\alpha_{s1}
 \; ,
 \end{split}
\end{equation}

In order to simplify the calculation, the same damping rate and the same classical pumping laser amplitudes in the cavity are assumed (i.e., $\gamma_{p1}=\gamma_{p2}=\gamma_{a}$, $\gamma_{i1}=\gamma_{s1}=\gamma_{b}$, $\gamma_{s2}=\gamma_{i2}=\gamma_{c}$  and $\epsilon_{1}=\epsilon_{2}=\epsilon$ ). Similarly, the same nonlinear coupling coefficient are assumed (i.e., $k_{2}=k_{3}$ ).

Threshold can be obtained if $k_{1}\geq 2k_{2} \sqrt{\gamma_{b}/\gamma_{c}}$. The values of the two threshold are expressed as following respecively,
\begin{equation}
\epsilon_{th}=\gamma_{a}\sqrt{\frac{k_{1}\gamma_{c}-\sqrt{k_{1}^{2}\gamma_{c}^{2}-4k_{2}^{2}\gamma_{b}\gamma_{c}}}{2k_{2}^{2}}} \;
\text{and} \; \;
\epsilon_{th}^{'}=\gamma_{a}\sqrt{\frac{k_{1}\gamma_{c}+\sqrt{k_{1}^{2}\gamma_{c}^{2}-4k_{2}^{2}\gamma_{b}\gamma_{c}}}{2k_{2}^{2}}} \; .
\end{equation}

When $\epsilon_{th}<\epsilon<\epsilon_{th}^{'}$, there is only one set of nonzero solutions for all six modes. If $\epsilon_{th}^{'}<\epsilon$, there will be two sets of nonzero solutions for all six modes. When there exists a threshold the analytical expressions of stationary solutions for different values of the pump amplitude are

(i) $\epsilon \leq \epsilon_{th}$
\begin{equation}
A_{i}=\frac {\epsilon}{\gamma_{a}} \; ,(i=p1,p2);\;A_{i}=0\;,(i=i1,s1,i2,s2)\;,
\end{equation}

(ii) $\epsilon_{th}<\epsilon \leq \epsilon_{th}^{'}$
\begin{equation}
 \begin{split}
A_{i}=A_{a}=\frac {\epsilon_{th}}{\gamma_{a}} \; &,(i=p1,p2);\\
A_{i}=\sqrt{\frac{\epsilon-\gamma_{a}A_{a}}{k_{1}A_{a}}}  \;,(i=i1,s1)&;\; A_{i}=\frac{k_{2}A_{a}^{2}A_{b}}{\gamma_{c}} \;,(i=i2,s2)\;,
 \end{split}
\end{equation}

(iii) $\epsilon > \epsilon_{th}^{'}$
\begin{equation}
 \begin{split}
A_{i}=A_{a}=\frac {\epsilon_{th}}{\gamma_{a}} \; \text{or} \; A_{i}=&A_{a}=\frac {\epsilon_{th}^{'}}{\gamma_{a}}\; ,\;(i=p1,p2);\\
A_{i}=\sqrt{\frac{\epsilon-\gamma_{a}A_{a}}{k_{1}A_{a}}}  \;,(i=i1,s1);&\; A_{i}=\frac{k_{2}A_{a}^{2}A_{b}}{\gamma_{c}} \;,(i=i2,s2)\;.
 \end{split}
\end{equation}

If $k_{1} < 2k_{2} \sqrt{\gamma_{b}/\gamma_{c}}$, there is no threshold in this system. On this condition the signal and the idler modes will not be excited no matter how large the pump amplitude is. Thus, when a threshold does not exist the analytical expressions of the stationary solutions are
\begin{equation}
A_{i}=\frac {\epsilon}{\gamma_{a}} \; ,(i=p1,p2);\;A_{i}=0\;,(i=i1,s1,i2,s2)\;.
\end{equation}

In the following, we decompose the system variables $\alpha_{i}$ into their steady-state values $A_{i}$ and small fluctuations around the steady-state values $\delta\alpha_{i}$ as $\alpha_{i}=A_{i}+\delta\alpha_{i}$.
Then equation (5) can be linearized as
\begin{equation}
\frac{\partial}{\partial t} \delta\tilde{\alpha} = -\mathbf{M}\delta\tilde{\alpha} +\mathbf{B\eta} \; ,
\end{equation}
where $\delta\tilde{\alpha}=[\delta\alpha_{p2},\delta\alpha_{p1},\delta\alpha_{i1},\delta\alpha_{s1},\delta\alpha_{i2},\delta\alpha_{s2},
\delta\alpha_{p2}^{*},\delta\alpha_{p1}^{*},\delta\alpha_{i1}^{*},\delta\alpha_{s1}^{*},\delta\alpha_{i2}^{*},\delta\alpha_{s2}^{*}]^{T}$;  M, the drift matrix, is given by

\begin{equation}
\mathbf{M} =
\left( \begin{array}{ccc}
\mathbf{m_{1}} & \mathbf{m_{2}} \\
\mathbf{m_{2}^{*}} & \mathbf{m_{1}^{*}}\\
\end{array} \right)\;
\end{equation}
where
\begin{equation}
\mathbf{m_{1}} =
\left( \begin{array}{cccccc}
\gamma_{a} & 0 & k_{1}A_{a}A_{b} &   k_{1}A_{a}A_{b}-k_{2}A_{a}A_{c} & k_{2}A_{a}A_{b} & 0 \\
0 & \gamma_{a} & k_{1}A_{a}A_{b}-k_{2}A_{a}A_{c} & k_{1}A_{a}A_{b} & 0 & k_{2}A_{a}A_{b} \\
-k_{1}A_{a}A_{b} & -k_{1}A_{a}A_{b}+k_{2}A_{a}A_{c} & \gamma_{b} & 0 & k_{2}A_{a}^{2} & 0\\
-k_{1}A_{a}A_{b}+k_{2}A_{a}A_{c} & -k_{1}A_{a}A_{b} & 0 & \gamma_{b} & 0 & k_{2}A_{a}^{2} \\
-k_{2}A_{a}A_{b} & 0 & -k_{2}A_{a}^{2} & 0 &  \gamma_{c} & 0 \\
0 & -k_{2}A_{a}A_{b} & 0 & -k_{2}A_{a}^{2} & 0 &  \gamma_{c} \\
\end{array} \right)\; ,
\end{equation}
and
\begin{equation}
\mathbf{m_{2}} =
\left( \begin{array}{cccccc}
0 & k_{1}A_{b}^{2} & k_{2}A_{a}A_{c} & 0 & 0 & -k_{2}A_{a}A_{b} \\
k_{1}A_{b}^{2} & 0 & 0 & k_{2}A_{a}A_{c} & -k_{2}A_{a}A_{b} & 0\\
k_{2}A_{a}A_{c} & 0 & 0 & -k_{1}A_{a}^{2} & 0 & 0\\
0 & k_{2}A_{a}A_{c} & -k_{1}A_{a}^{2} & 0 & 0 & 0\\
0 & -k_{2}A_{a}A_{b} &0 & 0 & 0 & 0\\
-k_{2}A_{a}A_{b} & 0 & 0 & 0 & 0 & 0\\
\end{array} \right)\; .
\end{equation}

If the requirement that the real part of the eigenvalues stay positive is satisfied, the fluctuation equations will describe an Ornstein-Uhlenbeck process \cite{Gardiner2002} for which the intracavity spectral correlation matrix is
\begin{equation}
\mathbf{S}(\omega) = (\mathbf{M}+i\omega\mathbf{I})^{-1} \mathbf{BB^{T}} (\mathbf{M^{T}}-i\omega\mathbf{I})^{-1}    \; .
\end{equation}

All the correlations required to study the measurable extracavity spectra are contained in this intracavity spectral matrix. The output fields can be obtained along with the standard input-output relations \cite{Gardiner1985}. In particular, the output fields spectral variances and covariances have the general form
\begin{equation}
 \begin{split}
S_{X_{i}}^{out}(\omega)=&1+2\gamma_{i}S_{X_{i}}(\omega)\; ,\\
S_{X_{i},X_{j}}^{out}(\omega)=&2\sqrt{\gamma_{i}\gamma_{j}}S_{X_{i},X_{j}}(\omega)\; .
 \end{split}
\end{equation}
Similar expressions can be derived for the $\hat{Y}$ quadratures.

\section{Multipartite Entanglement and Output Spectra}

In order to investigate multipartite entanglement, and particularly show that the system under consideration demonstrates true six-partite entanglement, we  define quadrature operators for each mode as
\begin{equation}
\hat{X}_{i} = \hat{a}_{i} + \hat{a}^{\dagger}_{i} \; ,\;\; \hat{Y}_{i} = -i( \hat{a}_{i} - \hat{a}^{\dagger}_{i} ) \; ,
\end{equation}
such that $[\hat{X}_{i},\hat{Y}_{i}]=2i$. The conditions proposed by van Loock and Furusawa \cite{vanloock2003}, which are a generalization of the conditions for bipartite entanglement, are sufficient to demonstrate multipartite entanglement.

Using the quadrature definitions, the six-partite inequalities are
\begin{equation}
 \begin{split}
V(X_{i2}-X_{p1})&+V(g_{p2}Y_{p2}+Y_{p1}+g_{i1}Y_{i1}+g_{s1}Y_{s1}+Y_{i2}+g_{s2}Y_{s2})\geq 4 \; ,\\
V(X_{p1}+X_{s1})&+V(g_{p2}Y_{p2}+Y_{p1}+g_{i1}Y_{i1}-Y_{s1}+g_{i2}Y_{i2}+g_{s2}Y_{s2})\geq 4 \; ,\\
V(X_{s1}-X_{i1})&+V(g_{p2}Y_{p2}+g_{p1}Y_{p1}+Y_{i1}+Y_{s1}+g_{i2}Y_{i2}+g_{s2}Y_{s2})\geq 4 \; ,\\
V(X_{i1}+X_{p2})&+V(-Y_{p2}+g_{p1}Y_{p1}+Y_{i1}+g_{s1}Y_{s1}+g_{i2}Y_{i2}+g_{s2}Y_{s2})\geq 4 \; ,\\
V(X_{p2}-X_{s2})&+V(Y_{p2}+g_{p1}Y_{p1}+g_{i1}Y_{i1}+g_{s1}Y_{s1}+g_{i2}Y_{i2}+Y_{s2})\geq 4 \; ,
 \end{split}
\end{equation}
where $V(\hat{A})=\langle\hat{A}^{2}\rangle-\langle\hat{A}\rangle^{2}$ denotes the variance and $g_{i}$ are arbitrary real parameters used to optimize the violation of these inequalities. It is important to note that in the uncorrelated limit these optimized VLF criteria approach 4. Hence, without optimization, some entanglement which is presented may be missed.


From now on, we numerically calculate the values of VLF inequalities, and these are the quantities that can be measured in experiments. Note that in our symmetrical situation, inequalities i2-p1 and s2-p2 are equal; the same are to p1-s1 and i1-p2. To simplify, we write the subscript s1-i1 as A, p1-s1 and i1-p2 as B, i2-p1 and s2-p2 as C separately.

When below the threshold or without threshold, the powers of the pump are much higher than others and the steady-state values of the other output beams are all equal to zero . Therefore, the quantum characteristics of the pump and other beams in such situation are not considered.

\begin{figure}
\begin{minipage}[t]{0.5\linewidth}
\centering
\includegraphics[width=\textwidth]{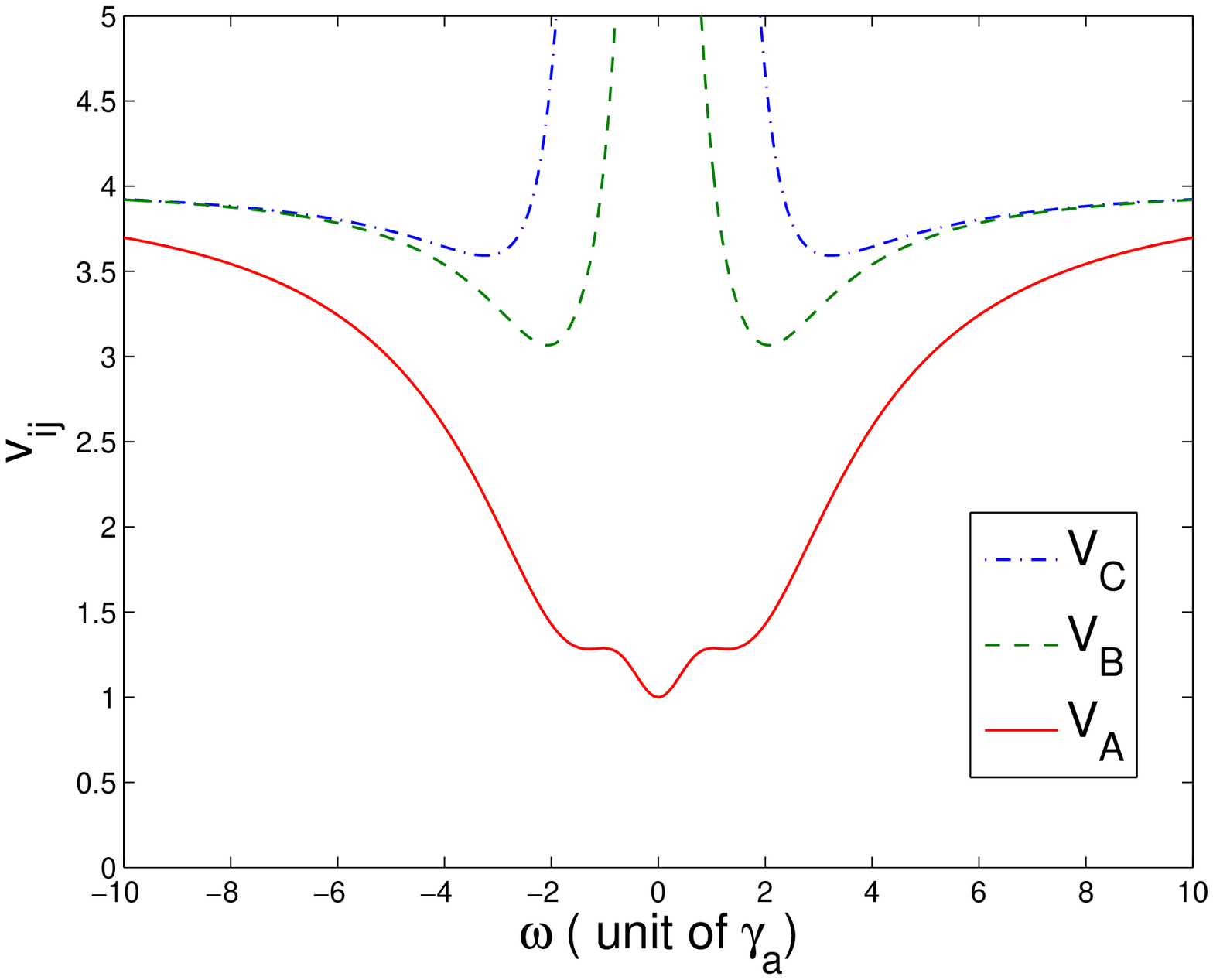}
\caption{(Color online) Minima of the inequalities as a \\function of the analysis frequency normalized to $\gamma_{a}$ with \\$\gamma_{a}=\gamma_{b}=\gamma_{c}=0.03 , k_{1}=2k_{2}=1$,  and $\epsilon=1.5\epsilon_{th}$. There \\is only one threshold in this case.}
\label{fig:side:a}
\end{minipage}%
\begin{minipage}[t]{0.5\linewidth}
\centering
\includegraphics[width=\textwidth]{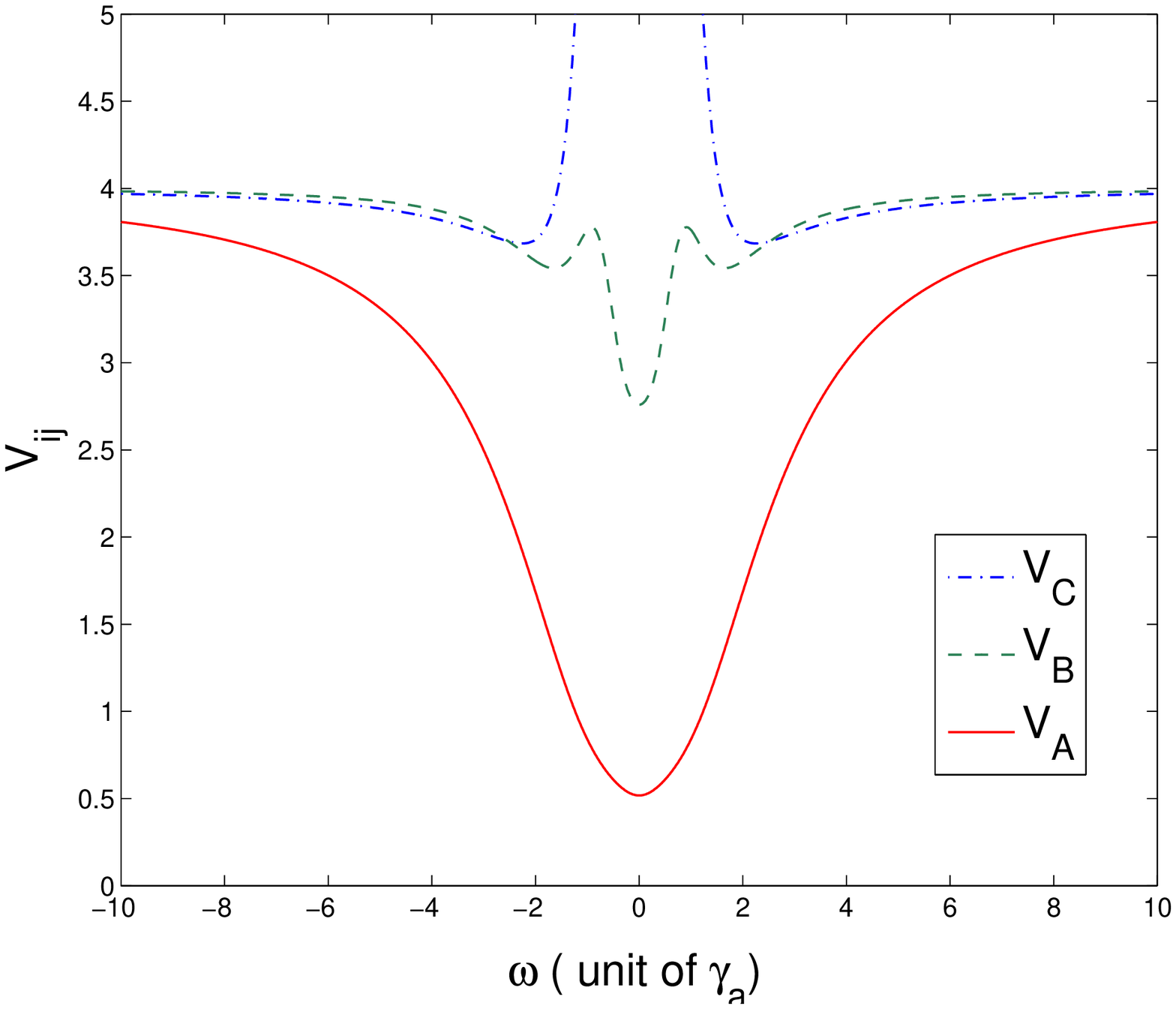}
\caption{(Color online) Minima of the inequalities as a \\function of the analysis frequency normalized to $\gamma_{a}$ with \\$\gamma_{a}=\gamma_{b}=\gamma_{c}=0.03 , k_{1}=1, k_{2}=0.4$,  and $\epsilon=1.2\epsilon_{th}$. \\Here $\epsilon_{th}^{'}=2\epsilon_{th}$ and $\epsilon_{th}<\epsilon \leq \epsilon_{th}^{'}$.}
\label{fig:side:b}
\end{minipage}
\end{figure}

In Fig 2, we plot the minimum of the inequalities as a function of frequency normalized to $\gamma_{a}$ with $\gamma_{a}=\gamma_{b}=\gamma_{c}=0.03 , k_{1}=2k_{2}=1$,  and $\epsilon=1.5\epsilon_{th}$. In this case, $k_{1} = 2k_{2} \sqrt{\gamma_{b}/\gamma_{c}}$ so that the two supposed thresholds coincide to be one threshold. It is obvious that the minimal values of inequalities are all less than 4 in a wide range of analysis frequencies, which is sufficient to demonstrate that all the six beams are CV entangled with each other. In general, the smaller the value of the inequality, the stronger the degree of entanglement and quantum correlation is. For large frequencies the inequalities approaches 4, which is the result of our optimization. In Fig. 2, we can also see that the value of $V_{A}$ is comparatively smaller than $V_{B}$, which indicates that the degree of entanglement of second-class output is bigger. It is similar to those situations in the four wave mixing process.

In Fig. 3, we plot the minimum of the inequalities as a function of frequency normalized to $\gamma_{a}$ with $k_{2}=0.4, \epsilon=1.2\epsilon_{th}$ and the other parameters remain unchanged. Again, the minima of the inequalities are less than 4 when above threshold, and it verifies that the all field modes are entangled.

\begin{figure}
\begin{minipage}[t]{0.5\linewidth}
\centering
\includegraphics[width=\textwidth]{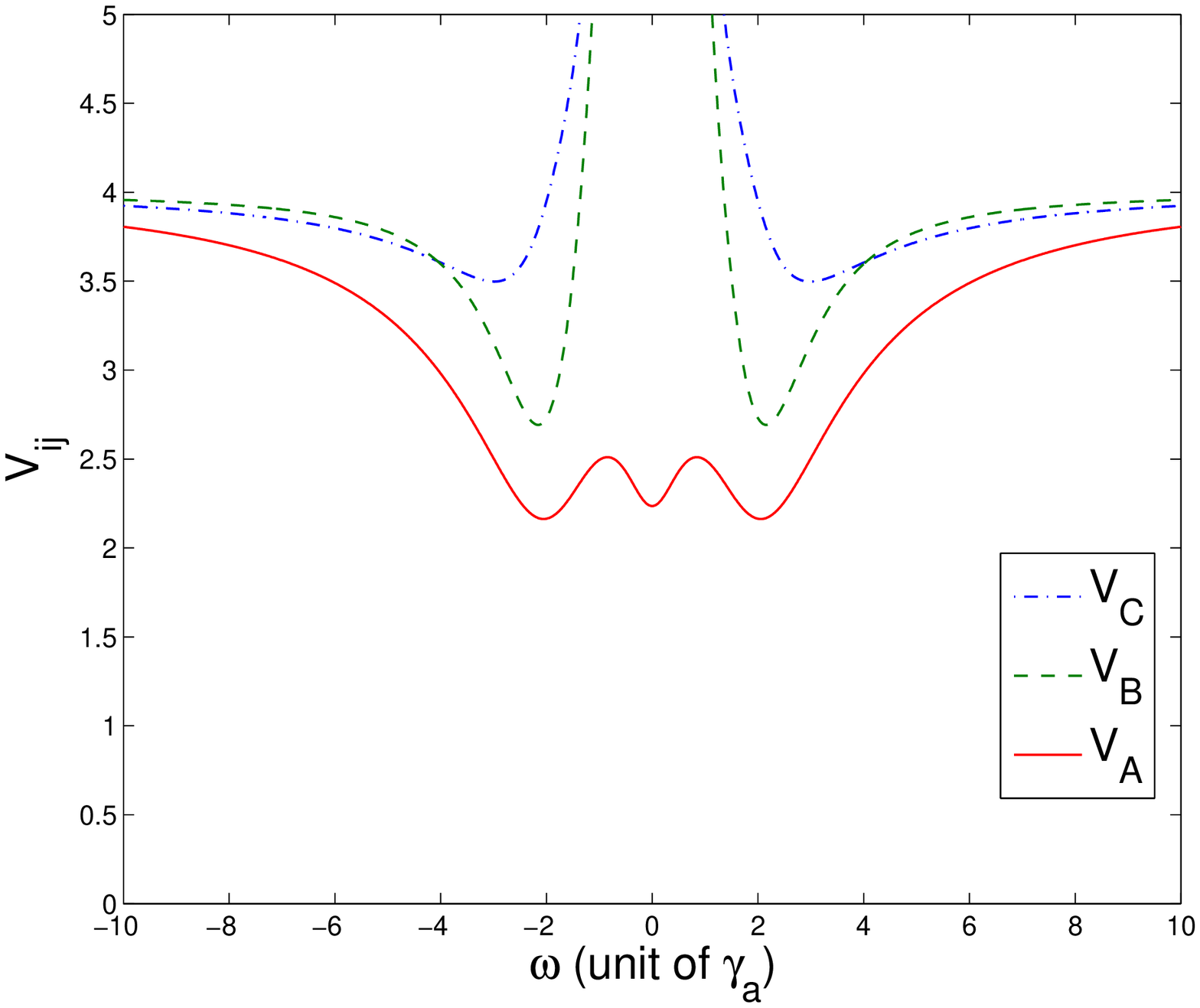}
\caption{(Color online) Minima of the inequalities as a \\function of the analysis frequency normalized to $\gamma_{a}$ with \\$\gamma_{a}=\gamma_{b}=\gamma_{c}=0.03 , k_{1}=1, k_{2}=0.4$,  and $\epsilon=1.1\epsilon_{th}^{'}=\\2.2\epsilon_{th}$. We use the stationary solution with $A_{a}=\epsilon_{th}/\gamma_{a}$.}
\label{fig:side:a}
\end{minipage}%
\begin{minipage}[t]{0.5\linewidth}
\centering
\includegraphics[width=\textwidth]{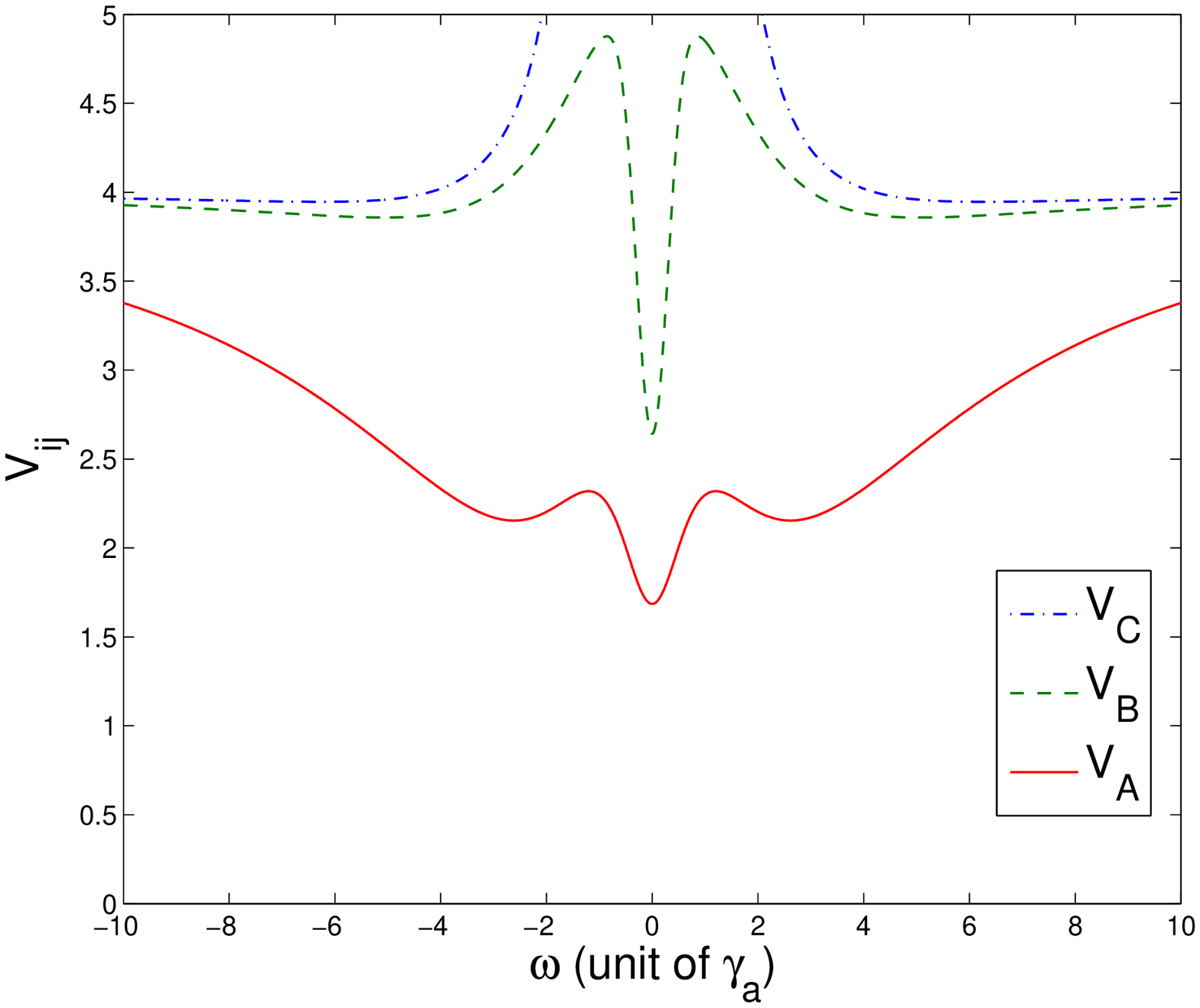}
\caption{(Color online) Minima of the inequalities as a \\function of the analysis frequency normalized to $\gamma_{a}$ with\\ the parameters unchanged as in Fig. 4. Here we use the\\ stationary solution with $A_{a}=\epsilon_{th}^{'}/\gamma_{a}$.}
\label{fig:side:b}
\end{minipage}
\end{figure}

However, we are unable to decide the real entanglement when $\epsilon > \epsilon_{th}^{'}$ because of the two possible stationary solutions. Because the solution of ordinary differential equations (ODE) usually depends on the initial value of variables. While the variables of ODE here are set initially as Gaussian states and randomly generated at each specific situation. It should be expected that all those variables will fall into one of the stationary solutions as the time approaches infinity, with a certain distribution of these two stationary solutions.  Therefore, we calculate the minimum of the inequalities beyond the larger threshold $\epsilon_{th}^{'}$ using each stationary solution separately.
\begin{figure}
\begin{minipage}[t]{0.5\linewidth}
\centering
\includegraphics[width=\textwidth]{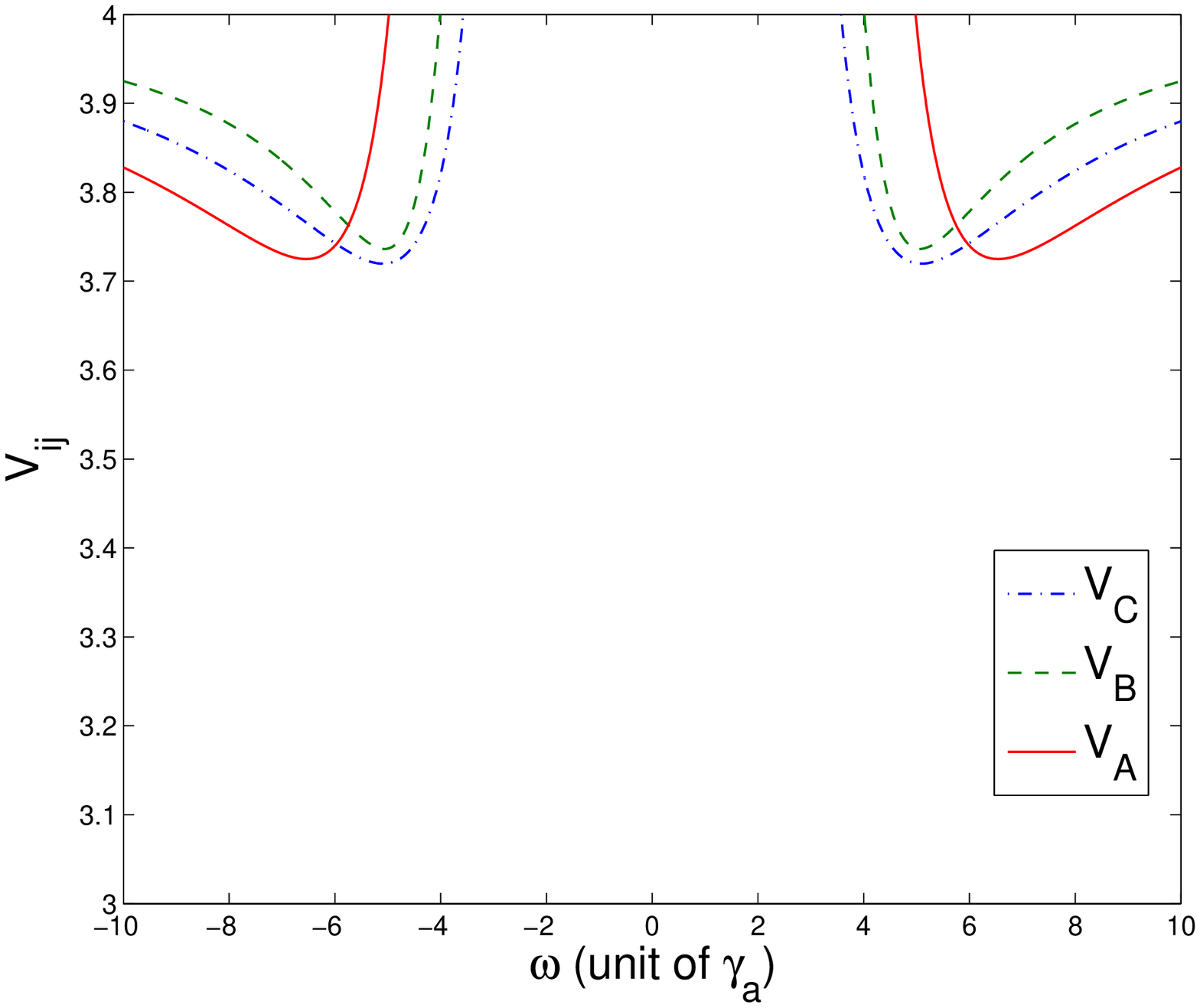}
\caption{(Color online) Minima of the inequalities as a \\function of the analysis frequency normalized to $\gamma_{a}$ with \\ $\epsilon=2.2\epsilon_{th}^{'}=4.4\epsilon_{th}$ and other parameters unchanged as \\in Fig. 4. We use the stationary solution with $A_{a}=\epsilon_{th}/\gamma_{a}$.}
\label{fig:side:a}
\end{minipage}%
\begin{minipage}[t]{0.5\linewidth}
\centering
\includegraphics[width=\textwidth]{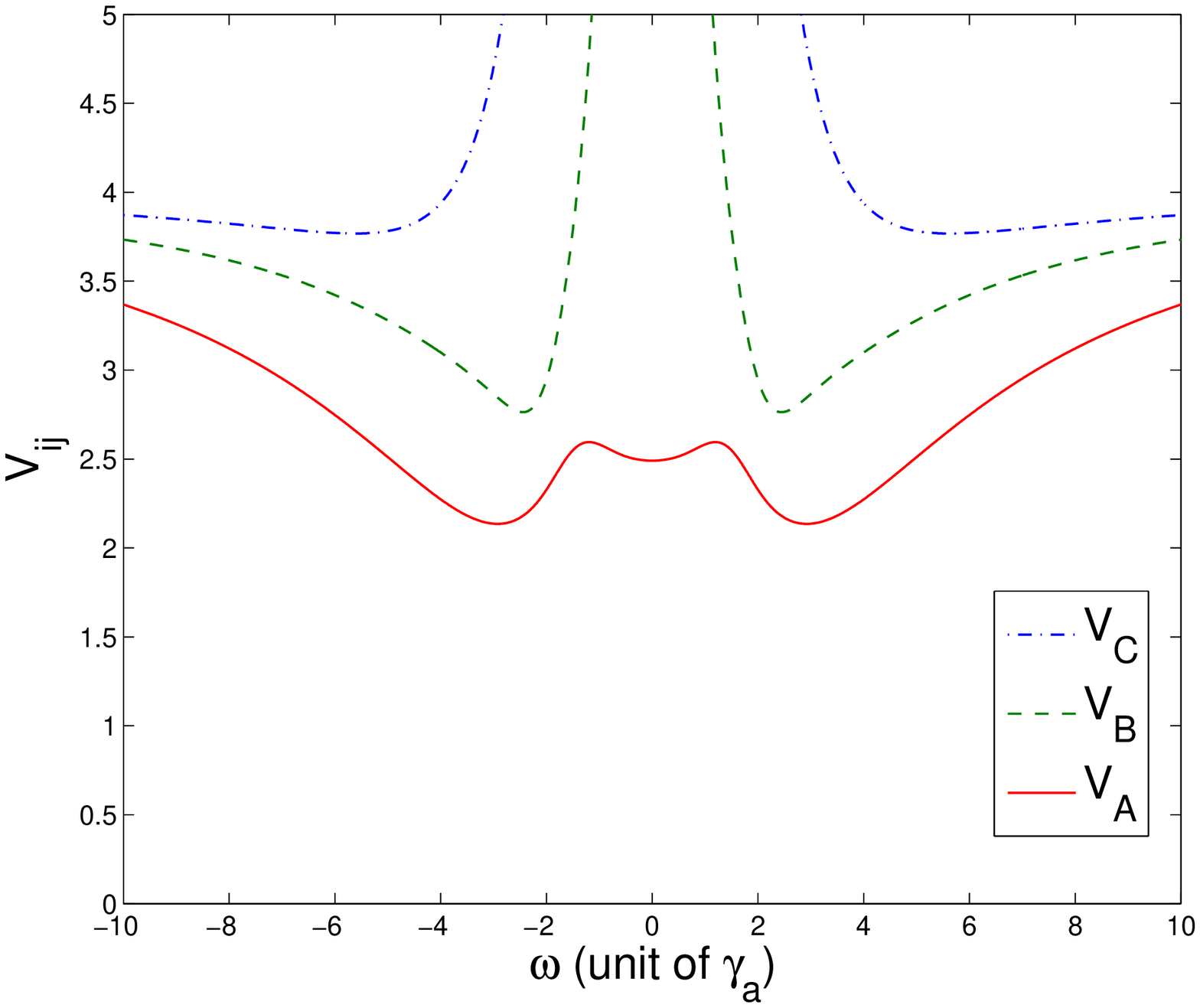}
\caption{(Color online) Minima of the inequalities as a \\function of the analysis frequency normalized to $\gamma_{a}$ with\\ the parameters unchanged as in Fig. 6. Here we use the\\ stationary solution with $A_{a}=\epsilon_{th}^{'}/\gamma_{a}$.}
\label{fig:side:b}
\end{minipage}
\end{figure}

In the following, we plot the minimum of the inequalities as a function of frequency normalized to $\gamma_{a}$  using each stationary solution at different $\epsilon$, with other parameters unchanged. We find an interesting result. As shown in Fig. 4 and Fig. 5, when $\epsilon$ is a little bigger than $\epsilon_{th}^{'}$, the stationary solution $A_{a}=\epsilon_{th}/\gamma_{a}$ shows a relatively competitive result, just as results in Fig. 2 and Fig. 3. The stationary solution $A_{a}=\epsilon_{th}^{'}/\gamma_{a}$ does not work effective and the minimum value of $V_{C}$ is almost near 4. It is because the output beams other than the pump is so lightly occupied if $A_{a}=\epsilon_{th}^{'}/\gamma_{a}$. In other words, $A_{i}$ (i=s1,i1,s2,i2) is close to zero.

On the other hand, in Fig. 6 and Fig. 7, the situation is reverse. For solution $A_{a}=\epsilon_{th}/\gamma_{a}$, it seems that, as the intensity of the pump power increases, the gain of other beams begins to saturate. Usually, when the intensity of the pump beams is higher than those of others, their quantum characteristics tend to vanish and the degree of quantum entanglement will be small. However the solution $A_{a}=\epsilon_{th}^{'}/\gamma_{a}$ is greater at this time, which seems to be more effective in a wide range of $\epsilon$.
\begin{figure}
\begin{minipage}[t]{0.5\linewidth}
\centering
\includegraphics[width=\textwidth]{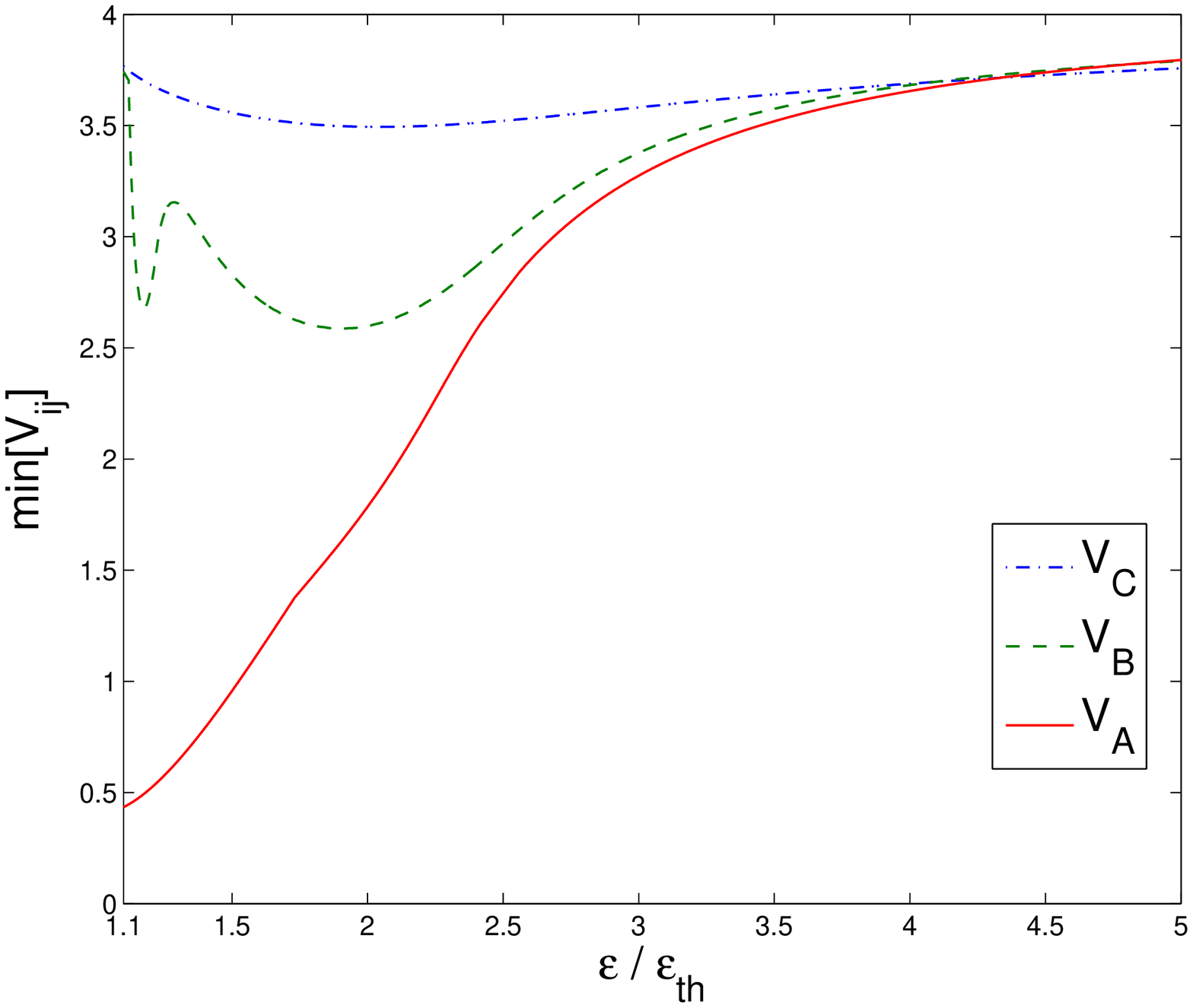}
\caption{(Color online) Maximum quadripartite entangle-\\ment as a function of the pump power parameter $\epsilon$
 with \\the parameters unchanged as in Fig. 4. Here we use the\\ stationary solution with $A_{a}=\epsilon_{th}/\gamma_{a}$.}
\label{fig:side:a}
\end{minipage}%
\begin{minipage}[t]{0.5\linewidth}
\centering
\includegraphics[width=\textwidth]{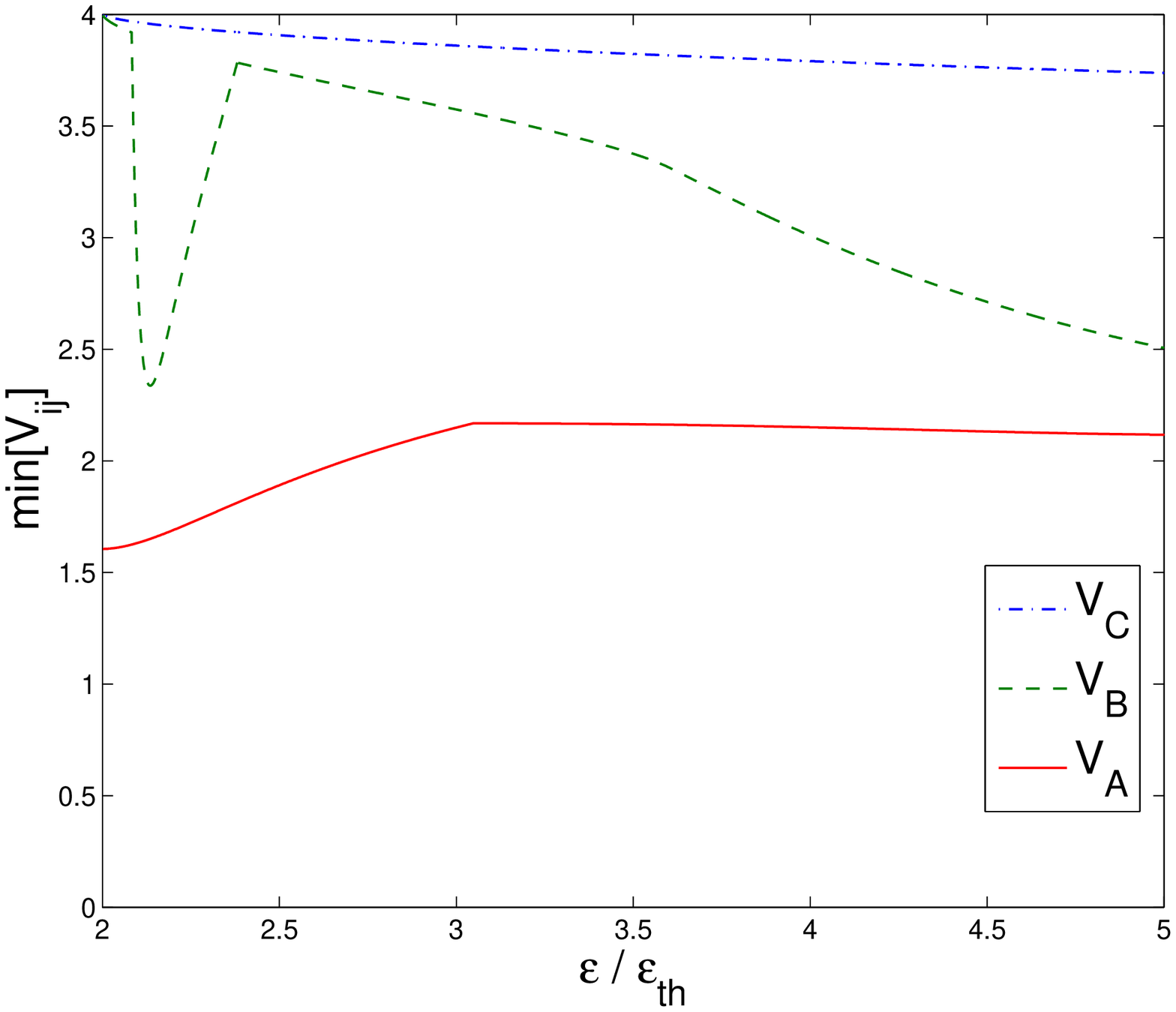}
\caption{(Color online) Maximum quadripartite entangle-\\ment as a function of the pump power parameter $\epsilon$
 with \\the parameters unchanged as in Fig. 4. Here we use the\\ stationary solution with $A_{a}=\epsilon_{th}^{'}/\gamma_{a}$.}
\label{fig:side:b}
\end{minipage}
\end{figure}

In Fig. 8 and Fig. 9, we numerically search the minimum of $V_{ij}$ in all frequencies, as a function of the pump power parameter $\epsilon$, also with other parameters unchanged. In these figures, we can see clearly that the stationary solution $A_{a}=\epsilon_{th}/\gamma_{a}$ is valid when $\epsilon$ is small just like the usual cases, see Ref. \cite{Yu2011}. And solution $A_{a}=\epsilon_{th}^{'}/\gamma_{a}$ is effective even $\epsilon$ is tens of times larger than $\epsilon_{th}^{'}$. This directly confirms previous idea.
We hope this can be considered as an effect of optical bistability. Although we cannot confirm the proportion of photons resulting in the stationary solution $A_{a}=\epsilon_{th}^{'}/\gamma_{a}$, it really enlarges the range of $\epsilon$ to be effective.

\section{Conclusions}
We propose a scheme to produce bright six-partite CV entanglement in an optical cavity operating above threshold in a four-level atomic system, which has been theoretically demonstrated using both VLF criterion of multipartite CV entanglement and the input-output relations. This result is experimentally feasible and can have a larger range of pump power to produce an effective multi-partite entanglement; it is much well-developed, compared with a basic system using the third-order nonlinearity media. Moreover, the bright six-partite entangled beams could be separated spatially, with narrow linewidths after generation, which have significant use in quantum network and free-space teleportation.

\section{Acknowledgements}
\vspace{0.2in}

This work is supported by the National Natural Science Foundation of China (Grants No. 601102053), and Shanghai Jiao Tong University PRP (Grants No. T030PRP19035).

\end{document}